\documentclass{article}
\usepackage{spconf,amsmath,graphicx}
\DeclareMathOperator*{\argmax}{arg\,max} 
\usepackage[ruled,lined]{algorithm2e}
\usepackage{booktabs}
\usepackage{float}
\usepackage{graphicx}

\title{Curriculum optimization for low-resource speech recognition}
\name{Author(s) Name(s)\thanks{Thanks to XYZ agency for funding.}}
\address{Author Affiliation(s)}
\name{Anastasia Kuznetsova,$^{\dagger \ddagger}$\sthanks{The work was partially completed while interning at Rev.com.}
Anurag Kumar,$^{\dagger\star}$  Jennifer Drexler Fox,$^{\star}$  Francis M. Tyers$^{\ddagger}$}
  
  \address{$^{\dagger}$ Department of Computer Science, Indiana University Bloomington, USA \\
  $^{\ddagger}$ Department of Linguistics, Indiana University Bloomington, USA\\
      $^{\star}$ Rev.com}
      
\begin{document}
%
\maketitle
\begin{abstract}
Modern end-to-end speech recognition models show astonishing results in transcribing audio signals into written text. However, conventional data feeding pipelines may be sub-optimal for low-resource speech recognition, which still remains a challenging task. We propose an automated curriculum learning approach to optimize the sequence of training examples based on both the progress of the model while training and prior knowledge about the difficulty of the training examples. We introduce a new difficulty measure called \textit{compression ratio} that can be used as a scoring function for raw audio in various noise conditions. The proposed method improves speech recognition Word Error Rate performance by up to 33\% relative over the baseline system\footnote{© 20XX IEEE. Personal use of this material is permitted. Permission from IEEE must be obtained for all other uses, in any current or future media, including reprinting/republishing this material for advertising or promotional purposes, creating new collective works, for resale or redistribution to servers or lists, or reuse of any copyrighted component of this work in other works.}. 

\end{abstract}
\section{Introduction}
\label{sec:intro}

Automatic speech recognition (ASR) is the task of transcribing audio signals into text. Modern end-to-end ASR systems have achieved significant performance boosts over the past decade using various architectures and feature representations. However, ASR for low-resource languages still remains challenging. Conventional approaches to low-resource ASR such as transfer learning \cite{Yi:2019}, multilingual ASR \cite{Cho:2018}  or meta-learning \cite{Xiao:2021} adhere to the notion of world languages sharing low-level components despite phonetic, structural and orthographic differences. Transfer learning uses the parameters of the model trained on a data-rich language to fine-tune the target language system, while multilingual ASR model is trained on several languages by sharing the hidden layer. Meta-learning addresses the problem of redundancy of language-specific features learned by multilingual models and further optimizes meta-loss to achieve language-independent acoustic representations. While these techniques improve WER, it is still much higher than in high-resource systems. More recent ASR approaches turned to self-supervised representations  to reduce the need of labelled data while relying on large corpora of unlabelled audio \cite{Schneider:2019}. We propose an alternative approach achieving high recognition quality. Our model accounts for both the lack of training data and  diverse noise/quality conditions based on the idea of \textit{curriculum learning}.


\textit{Curriculum learning} is a paradigm in supervised machine learning first introduced by \cite{Bengio:2009}.  
It implies imitation of human learning strategies where the \textit{curriculum} is a set of tasks organised in order of increasing complexity. Early curriculum learning strategies used a ranking function to determine the complexity of the data and organised the sequence of training examples according to this measure \cite{Bengio:2009}. This is analogous to a teacher designed curriculum based on prior knowledge about the data, but it does not reflect the competence of the student over time.  \cite{Jiang:2015, Zhou:2018, Hacohen:2019,  Zhou:2020} show that a curriculum yields better results with adaptive learning that accounts for the history of acquired competence during the previous time steps. \cite{Zhou:2020} studied the dynamics of deep neural networks trained with dynamic instance hardness and showed that the model revisits harder samples more often due to higher variance in gradient values while easier examples tend to stay in the minima as soon as the minima are reached. However, \cite{Hacohen:2019} finds that in different empirical settings both learning harder and easy tasks first can benefit the model. 

Curriculum learning has been successfully used in natural language processing tasks such as language modelling \cite{Bengio:2009, Graves:2017} neural machine translation (NMT) \cite{Wang:2018, Platanios:2019, Liu:2020}, keyword spotting \cite{Takuya:2021}, and speech recognition \cite{Braun:2017, Suyoun:2017}. Most commonly two complexity strategies are employed: model competence-based \cite{Wang:2018, Platanios:2019, Zhou:2020} and data-driven. Complexity measures for data-driven learning include sentence/utterance length \cite{Suyoun:2017}, language model score, \(n\)-gram size \cite{Bengio:2009, Graves:2017}, word frequency ranking, and sentence norm \cite{Liu:2020}. Speech processing systems can rely on speech to noise ratio (SNR) as a measure of difficulty by gradually blending more and more noise into clean speech signals \cite{Braun:2017, Takuya:2021}, but only if clean speech signals are available for creating noisy mixtures with different SNRs.

In this work, we propose a new complexity metric called \textit{compression ratio} that helps to rank training examples based on audio quality even with the absence of clean data. Additionally, we demonstrate that both an external teacher curriculum and learner's progress are important for low-resource ASR. The teacher curriculum acts as a reliable prior while the student can construct its own curriculum in an automated manner based on progress gains.  We experiment with compression ratio and text-based difficulty measures to  show that the signal-based prior leads to a more optimal solution. 

The remainder of the paper as organized as follows: Section \ref{sec:method} describes the complexity measures and methods; Sections \ref{sec:experiments} and \ref{sec:results} discuss the experimental setting and the results; Section \ref{sec:conclusion} summarizes our contributions.

\section{Method}
\label{sec:method}

Curriculum learning for neural network optimization has been implemented in a variety of ways. Approaches in \cite{Wang:2018, Hacohen:2019, Zhou:2020} propose a sampling strategy where a \textit{scoring} function is the source of prior knowledge for the model obtained either from the properties of the data or from a pre-trained `teacher' model and a \textit{pacing} function that adjusts the sampling weights depending on the student's progress. Our approach employs the \textit{reinforcement learning} paradigm adopted by \cite{Graves:2017, Kumar:2019}. Graves et al. \cite{Graves:2017} experiment with a k-armed bandit algorithm while \cite{Kumar:2019} use more complex deep q-learning networks. 

The reinforcement learning framework has three main concepts: \textit{agent}, \textit{environment} and \textit{reward}. The \textit{agent} is an entity residing in an environment; it selects the next best action following its \textit{policy} (a value function over all possible actions). The \textit{environment} generates the feedback reflecting the optimality of the current action. The environmental feedback is converted into a \textit{reward} -- a signal interpretable by the agent -- and the agent adjusts its policy based on the reward in order to take better actions in the future. Formulating curriculum generation in terms of reinforcement learning allows us to incorporate both  knowledge about the data and knowledge about the model's learning progress. Sections \ref{sub:complexity} and \ref{sub:curriculum} give a detailed description of curriculum formulation, task and complexity metric definitions.

\subsection{Complexity metrics}
\label{sub:complexity}

ASR systems receive a set of paired audio and text sequences as input. We can leverage both signal and text to build a scoring function. We hypothesize that, for speech recognition, acoustic properties have higher importance in guiding the neural network towards the optimum than textual features. Thus, we propose the \textit{compression ratio} metric to score ASR input signals. 

\begin{equation}
    \mathrm{CR} = 1 - \frac{\mathrm{Size}_{\mathrm{after}}}{\mathrm{Size}_{\mathrm{before}}}
    \label{eq:compression}
\end{equation}
Compression ratio is defined in  (\ref{eq:compression}) where $\mathrm{Size}_{\mathrm{before}}$ is size of audio file before the compression and $\mathrm{Size}_{\mathrm{after}}$ size after applying \textit{gzip} compression algorithm respectively.

\begin{figure}
    \centering
    \includegraphics[width=0.5\textwidth]{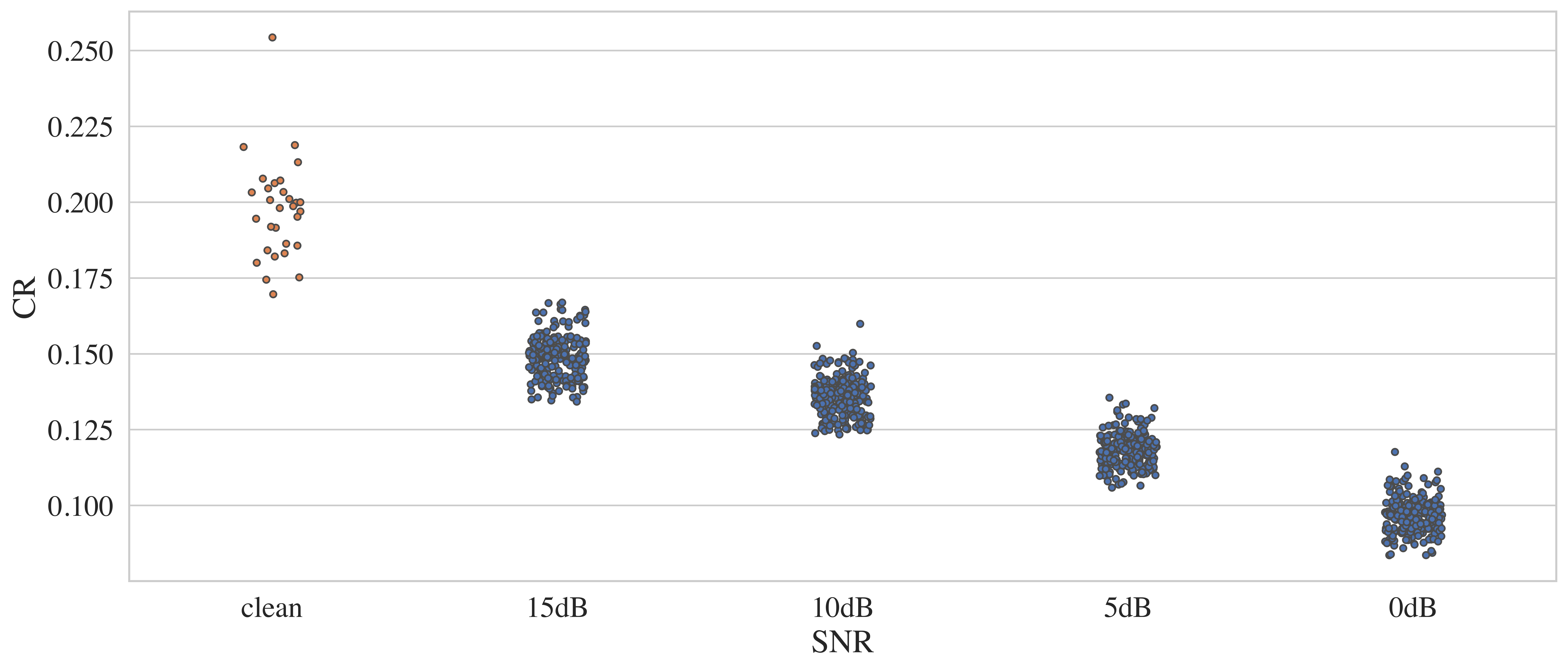}
    \caption{Compression ratio of clean audio compared to noisy mixtures at different SNRs (data from NOIZEUS\cite{Hu:2007}).}
    \label{fig:snr}
\end{figure}

We assume that if there is less noise in the signal, the entropy of the signal will be lower and the audio will be more compressed with respect to its uncompressed version. On the other hand, noisier signals will be harder to compress and will yield lower compression scores. Figure~\ref{fig:snr} demonstrates this and shows that clean signals are compressed more as opposed to noisy mixtures of different SNR levels.

We also experiment with text based metrics such as the commonly used \textit{sentence length} and \textit{sentence norm}. Sentence norm is obtained from the vector representation of a sentence. Liu \cite{Liu:2020} notes that word vector norm is influenced by the properties of the word e.g. it increases with a decrease of the
word frequency. Their experiments show that norm can be a good difficulty measure. We define sentence norm in (\ref{eq:snorm}) where $y_j$ is a Word2Vec 
representation of BPE segment obtained from the \textit{Sentencepiece} unigram model \cite{Kudo:2018} and $N$ is the length of the sentence.

\begin{equation}
    \text{SN} = \left\|\frac{1}{N} \Sigma_{j=1}^N (y_j) \right\|_2
    \label{eq:snorm}
\end{equation}
Given the selected difficulty criteria,  the \textit{curriculum} is a set of training examples sorted in descending order by hardness and split into $K$  subsets of data that we call \textit{tasks}. Thus, we define a task set  $D = \{D_{1}, D_{2}, D_{3}, ... D_{K}\}$ where $k$ is the index of the task and $K$ is the empirically derived total number of tasks. Each $D_k$ contains an equal number of training batches $\mathcal{B}_{\mathrm{bsize}}$ with the \(\mathrm{bsize}\) (mini-batch size) of input audio.

\subsection{Curriculum Learning Framework} 
\label{sub:curriculum}

\begin{algorithm}
\DontPrintSemicolon
\SetKwInOut{Input}{Input}
\textbf{Initialize:} $D = f(X)$,  $\pi \leftarrow 0$; \;

\Begin{
\For{$t\rightarrow T$}{
    Draw $k$ based on current $\pi$;\;
    $\mathcal{B}_{t, k}\leftarrow sample(D_k)$;\; 
    Train the model on $\mathcal{B}_{t, k}$;\;
    Observe progress gain $\nu_{SPG}$;\;
    $r_t \leftarrow  g(\nu_{SPG})$;\; 
    Update $\pi$ on $r_t$;
    }
}           
\caption{Curriculum Learning}
\label{alg:training}
\end{algorithm}

Our agent is defined as a $K$-armed bandit acting in the reward space $\mathcal{R}$ which aims to collect the maximum expected reward in the finite number of training steps. At each time step $t$ based on the expected payoff the agent selects the best action which in our definition is equivalent to task index $k$. After selecting $D_k$ the agent observes the reward $r_t\in \mathcal{R}$ and updates the corresponding action value in policy $\pi(k)$. 

The reward $r_t$ is calculated based on the progress signals from the network termed \textit{prediction gains}. We chose to experiment with \textit{self-prediction gain}, a loss-driven progress signal proposed by \cite{Graves:2017}. SPG is an unbiased estimate w.r.t. $D_k$, it calculates the gain on a batch $\mathcal{B}'$ sampled from the same task as the training batch $\mathcal{B}$ (\ref{eq:spg}).
\begin{equation}
     \nu_\mathrm{SPG} = L(\mathcal{B}', \theta) - L(\mathcal{B}', \theta') \qquad \mathcal{B}'\sim D_k
\label{eq:spg}
\end{equation}

Among the variety of \(K\)-armed bandit algorithms to optimize the curriculum we chose to experiment with stochastic \textit{EXP3.S} algorithm proposed by \cite{Auer:2002a} and tested on NLP tasks by \cite{Graves:2017}, and the deterministic \textit{SW-UCB\#} algorithm by \cite{Wei:2018}. As both the environment and parameters of the ASR model are changing, the best task with the maximum payoff may change over time. Both of these algorithms are adapted for non-stationary problems; they ensure a balance between exploitation (choosing the best $k$ at all times) and exploration (occasionally selecting a random $k$) and minimize the regret in the long run. EXP3.S uses the $\epsilon$-greedy strategy to sample random $k$ with $\epsilon$ probability and select the best task with probability $1-\epsilon$. The equation for policy update in EXP3.S is shown in (\ref{eq:policy_exp3}), where $K$ is the total number of tasks, $t$ is the current time step and $w$ is the weight for current $t$ and $k$.
\begin{equation}
    \pi_{t}^{\textsc{EXP3.S}} = (1-\epsilon) \frac{e^{w_{t, k}}}{\Sigma_{j=1}^K e^{w_{t, j}}} + \frac{\epsilon}{K}
\label{eq:policy_exp3}
\end{equation}
\noindent
On the other hand, arm selection in SW-UCB\# is based on the mean reward $\Bar{r}_k$ collected within a sliding window and the arm count within this time window. 
\begin{equation}
k_t = \argmax\{\Bar{r}_k(t-1, \alpha)+c_k(t-1 , \alpha)\}
\label{eq:ucb_arm}
\end{equation}

See (\ref{eq:ucb_arm}), where $\Bar{r}_k$ is the estimate of mean reward for arm $k$ within the sliding window, $n_k$ is the number of times arm $k$ was selected until time step $t$ and $\alpha$ controls the window size, $ c_k = \sqrt{\frac{(1+\alpha) \ln{t}}{n_k(t, \alpha)}}$. In deep neural network training, the magnitude of the loss is higher in the initial stages of training and gradually decreases towards the convergence. To account for this, SW-UCB\# has two environment modes, \textit{abruptly-changing} and \textit{slowly-varying}, with the mode determined by the variance of the recent loss history. $\alpha$, and thus the window size, is calculated differently in the two modes, so that smaller windows are used in the abruptly-changing mode. 

Algorithm~\ref{alg:training} shows a generalized version of Curriculum Learning. First, training data $X$ is scored according to complexity function $f(X)$. Weight vector $w$ and policy vector $\pi$ (both of size $K$) are initialized to \(0\). For each time step $t$, we select the best $k$ based on $\pi$, then we sample batch $\mathcal{B}_{t, k}$ from the corresponding task, train the model and observe the progress gain. Reward can be calculated directly from $\nu_{SPG}$ or mapped by $g(\cdot)$ to be in an interval $r_t\in[-1, 1]$ to avoid the effect of high loss magnitude \cite{Graves:2017}.

\section{Experiments}
\label{sec:experiments}

We evaluate the proposed approach on five low-resource languages from the multi-lingual crowed-sourced Common Voice 7.0 data set  \cite{Ardila:2020}: Basque (Eu), Frisian (Fy), Tatar (Tt),  Kyrgyz (Ky) and Chuvash (Cv). The amount of training portion of the data ranges from less than 2 hours of audio (Cv) to 45 hours (Eu). Table~\ref{tab:data} shows the statistics of the selected datasets.


\begin{table}
\centering
\scalebox{0.97}{
\begin{tabular}{lrrrrr}
\toprule
               & \textbf{Eu} & \textbf{Fy} & \textbf{Ky} & \textbf{Tt} & \textbf{Cv} \\
               \midrule
\textbf{Train} & 45:08     & 18:25      & 24:17      & 19:27    & 1:41     \\
\textbf{Dev}   & 8:03      & 4:14      & 2:07      & 2:47      & 0:45      \\
\textbf{Test}  & 8:34      & 4:25      & 2:12      & 5:07      & 1:08     \\
\bottomrule
\end{tabular}
}
\caption{Statistics for the datasets in hours and minutes.}
\label{tab:data}
\end{table}

\subsection{Experimental setup}
\label{sub:setup}
All models were trained in ESPNet using the hybrid CTC~/Attention model defined in the Common Voice recipe \cite{Watanabe:2018}. From the raw audio we extract 80-dimensional log-mel filter banks with a window length of 25ms and stride of 10ms. The data is then augmented using SpecAugment. 
We use Sentencepiece \cite{Kudo:2018} to extract BPE segments from the transcriptions for RNN LM training. The acoustic model is comprised of an encoder that has 9 Conformer 
blocks with 4 attention heads and a decoder that has 6 identical transformer blocks. 
The network is trained with Adam optimizer with the warmup phase of 25k steps. 
In order to achieve better performance on low-resource data we apply transfer learning from an English ASR model pretrained on Common Voice 7.0 English corpus with a Word Error Rate (WER) of 15.2. Evaluation is done by averaging the 10 best checkpoints based on validation accuracy. We report Word Error Rate (WER) and  Character Error Rate (CER).

\begin{figure*}[ht]
    \centering
    \includegraphics[width=0.99\textwidth]{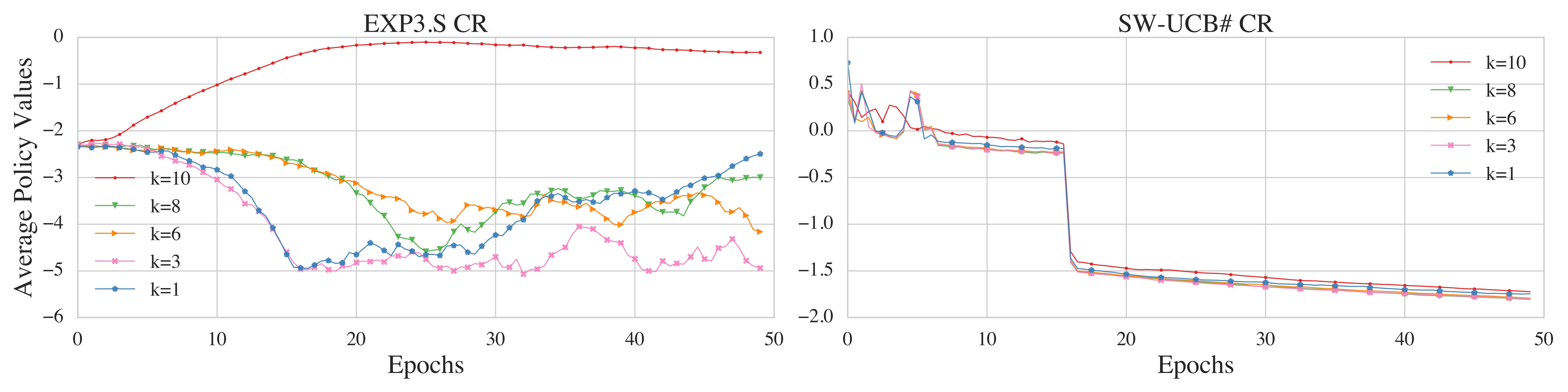}
    \caption{The figure shows log policy value change over time for selected $k$. This policy is generated by EXP3.S + CR and SW-UCB\# combinations on Kyrgyz (Ky) dataset and the policy is averaged per epoch. The highest $k=10$ has larger value throughout the training while other $k$ values are consistently lower.}
    \label{fig:policy}
\end{figure*}

\begin{table*}[ht!]
\centering
\scalebox{0.97}{
\begin{tabular}{lrrrrrrrrrr}
\toprule
               & \multicolumn{5}{c}{\textbf{WER}}                                    & \multicolumn{5}{c}{\textbf{CER}}                                                 \\
\midrule
\textbf{Model} & \textbf{Cv} & \textbf{Fy} & \textbf{Tt} & \textbf{Ky} & \textbf{Eu} & \textbf{Cv} & \textbf{Fy} & \textbf{Tt} & \textbf{Ky} & \textbf{Eu}  \\
\midrule
ESPNet + Trans & 61.4   & 9.6  & 23.5  & 5.5 & 7.5 & 15.9  & 3.1 & 5.9  & 2.3         & 1.5      \\
\midrule
EXP3.S + CR    & \textbf{41.8}       &   \textbf{7.8}          & \textbf{22.3}        & \textbf{4.1}         & 7.5         & \textbf{9.8}        &      \textbf{2.6}       & \textbf{5.6}         & \textbf{1.9}         & 1.5    \\
EXP3.S + SL    & \underline{\textbf{41.1}}        &    \textbf{9.3}         &      27.9       &   6.4          &      9.3       & \underline{\textbf{9.7}}         &       \textbf{3.3}      &      7.4       &  3.0           &    2.0       \\
EXP3.S + SN    & \textbf{42.7}       & \textbf{7.8}         &     24.2        &     \textbf{4.8}        &    8.1         & \textbf{10.2}        & \underline{\textbf{2.5}}         &       6.2      &     \textbf{2.2}        &      1.7 \\
SW-UCB\# + CR    & \textbf{42.7}       & \underline{\textbf{7.5}}         & \underline{\textbf{22.1}}        & \underline{\textbf{4.0}}        &      8.0       & \textbf{10.0}        & \underline{\textbf{2.5}}         & \underline{\textbf{5.4}}        & \underline{\textbf{1.8}}         &        1.6       \\
SW-UCB\# + SL    & \textbf{42.4}        &    10.9         & 24.0        &      6.7       &   10.0          & \textbf{9.9}         &     3.6        & 6.3         &    3.0         &        2.2      \\
SW-UCB\# + SN    & \textbf{42.6}        & \textbf{8.6}         & 24.8        &    \textbf{5.3}       &    8.6         & \textbf{10.1}        & \textbf{2.8}         & 6.4         &     \textbf{2.2}        &      1.8       \\
\bottomrule
\end{tabular}
}
\caption{The table shows WER and CER for five selected languages. Results in bold indicate the improvement over the baseline, underlined values indicate best result overall. Baseline results with bare transfer learning are shown in the first row. The results below show the combinations of the algorithm and complexity metric, \textit{CR} -- Compression Ratio, \textit{SL} -- Sentence Length, \textit{SN} -- Sentence Norm for $K=10$.}
\label{tab:results}
\end{table*}

In order to promote the most effective data usage we sample the audios with the highest policy values without replacement.
This ensures data variability and  better generalization of the model. Once the curriculum task with the highest value is exhausted, the training is continued using the rest of the data in order of decreasing policy values. The parameters used for SW-UCB\# are $\lambda=12$, $\gamma=0.4$, $\kappa=0.8$, $\nu=0.1$ and for EXP3.S $\epsilon=0.05$, $\eta=0.001$, $\beta=0$ and a history size of $10^3$ iterations.

\section{Results}
\label{sec:results}
We compare our approach to a transfer learning baseline trained with the parameters described in 
§~\ref{sub:setup}. The results are presented in Table~\ref{tab:results}.
Curriculum learning improves performance of the baseline ASR system with maximum $\downarrow$WER decrease of 33\% for Cv and minimum improvement of 5\% $\downarrow$WER for the Tt dataset. For Fy, Tt and Ky the best results are delivered by compression ratio metric. For Cv sentence length was more successful, however, the next lowest $\downarrow$WER is achieved with the combination of EXP3.S and compression ratio. The proposed approach gave better results in a very low data setting but did not improve on the larger Eu corpus.

Figure \ref{fig:policy} shows the policy derived by the EXP3.S and SW-UCB\# algorithms on Ky corpus. Policy values for EXP3.S start at the uniform distribution and then the hardest task $k=~10$ gets the highest policy value in the early stages of training. It means that harder $k$ is preferred earlier in the epoch and the model gets more information from the harder task. Task $k=1$ gradually increases the value towards the end of the training and the model selects easier batches in early iterations of final epochs.  In early training during the abruptly-changing phase SW-UCB\# switches between hard and easy task and then it follows the same pattern as EXP3.S when the higher $k$ is preferred continuing into the slowly-varying phase. The significant drop in SW-UCB\# policy values happens when the training mode is changing from abruptly- to slowly-varying. This is explained by  decreased reward and lower prediction gain magnitude, however this does not compromise the performance of the algorithm as $\argmax(\pi)$ of the policy is taken as opposed to sampling from the distribution in EXP3.S. 

\section{Conclusion}
\label{sec:conclusion}
 Curriculum learning yields decreased WER in 4 low-resource languages. Our approach has shown that the derived curriculum based on the student's learning progress improves performance and that the appropriate choice of complexity metric serves as a good prior for the model. Although text-based difficulty measures improve over the baseline model, CR outperforms them in several instances.  We believe that the performance of curriculum learning is dependent on the complexity metric distribution and can be further optimized by selecting other values of $K$. We leave this study for the future work.

\bibliographystyle{IEEEbib}
\bibliography{refs}

\end{document}